\documentclass[12pt]{iopart}
\usepackage{graphicx}
\begin{document}

\title[J. Phys.: Condens. Matter 18 (2006)]{Electronic properties of silica nanowires}
\author{G Bilalbegovi{\'c}}
\address{Department of Physics, University of Rijeka, Omladinska 14, 51000 Rijeka, Croatia}
\ead{goranka.bilalbegovic@zg.t-com.hr}

\begin{abstract}
Thin nanowires of silicon oxide were studied by pseudopotential density functional  electronic structure calculations
using the generalized gradient  approximation.
Infinite linear and zigzag Si-O chains were investigated.
A wire composed of three-dimensional periodically repeated Si$_4$O$_8$ units was also optimized, but this structure was found to be of limited stability.
The geometry, electronic structure, and Hirshfeld charges of these silicon oxide nanowires were computed.
The results show that the Si-O chain is metallic, whereas the zigzag chain and the Si$_4$O$_8$ nanowire are insulators.
\end{abstract}


\maketitle

\section{Introduction}
\label{sec:intro}

Nanochains of metals \cite{Ruitenbeek}, as well as of carbon, semiconductors and organic materials \cite{Ciraci,Service}
have  recently been the subject of
experimental and theoretical studies.
Similar chains of many other
chemical elements and compounds have not been studied. Because of
the present interest in nanotechnology these studies are
important. Chains with particular properties are candidates for preparation of nanostructures
with chosen applications. It is also possible to deposit chains on various substrates
and to obtain one-dimensional conductors and quantum confinement.

There are many crystalline phases of bulk silica, for example, quartz, tridymite, cristobalite, keatite, coesite,
and stishovite \cite{Wyckoff}.
In addition, amorphous SiO$_2$, which is abundant in nature, has
also been investigated and used in various technological applications.
These silica bulk phases have been studied by several experimental
\cite{Schnatterly,DiStefano}
and theoretical
\cite{Chelikowsky1,YXu,Allan,Charlier,Lee,Detraux,Donadio}
methods. Silica is a very good electrical insulator.
Macroscopic silica wires are used as waveguides in the visible and near-infrared spectral ranges.
Silica films are often applied in optics, and are used as electric and thermal insulators in electrical devices
\cite{Miyata}.
SiO$_2$ substrates are important in microelectronics, optics and  chemical applications. Therefore silica surfaces
have been also investigated
\cite{Rignanese}.

Much less study has been devoted to silica nanostructures.
Various cylindrical nanostructures of silica have recently been  synthesized:
nanowires, nanotubes, nanoflowers, bundles, and brush-like arrays
\cite{JLGole,DPYu,MZhang,Mazur}.
Their structural, mechanical, optical, and catalytic properties have been examined.
Silica nanowires, with diameters ranging from ten to several hundred nanometers, have been produced
using various experimental techniques. They  have been proposed for use as high-intensity light sources, near-field optical microscopy probes, and interconnections in integrated optical devices.

The properties of infinite silica chains have not been theoretically investigated.
However, several theoretical studies of silica clusters have been carried out
using Gaussian
\cite{LSWang,Nayak,Bromley,Sun,Bromley2,Jena,QWang},
GAMESS
\cite{WCLu},
SIESTA
\cite{Zhao}
and
DMOL
\cite{Nayak}
packages, as well as several other density functional theory  (DFT) programs
\cite{Chelikowsky,Chu,Song,Flikkema}.
Nanotubes of SiO$_x$, $x=1,2$, have recently been  studied using the VASP DFT program
\cite{Singh}.
All these computational studies of silica nanostructures,
have shown that their properties are often different in relation to the
bulk. Therefore, it is also important to study infinite
silica nanochains where periodic boundary conditions are used along
the axis. These one-dimensional structures of silica are
interesting from the theoretical point of view, as well as models
of very long real nanowires. They provide additional systems for
investigating the structure and bonding in silica materials, and
offer possibilities of designing  new nanostructures.
It is possible to prepare such thin silica wires on the substrates. The one that is
the most interesting for applications is the assembly of silica chains
on silicon surfaces and nanowires.

In this work, the
structure, energetics and electronic properties of thin
silica nanowires were investigated using a computational method. Infinite linear and zigzag chains, as well as a nanowire composed of periodically repeated Si$_4$O$_8$ structural units, were
constructed and optimized using a plane wave pseudopotential
approach to the density functional theory. The rest of the paper is
organized as follows. Section \ref{sec:2}  presents the method. In
section \ref{sec:3}  the results and discussion are given. Conclusions
are outlined in section \ref{sec:4}.

\section{Computational Method}
\label{sec:2}

\textit{Ab initio} DFT calculations \cite{Kohn,Sham} within the
plane-wave pseudopotential method were performed to study silica
chains. The pseudopotential approach has been very successful in
describing  the  structural and electronic properties of various
materials \cite{Martin}. The ABINIT code was used
\cite{Abinit}. The same method has already been applied to calculate
various properties of  bulk silica
\cite{Allan,Charlier,Lee,Detraux,Donadio} and the (0001)
$\alpha$-quartz surface \cite{Rignanese}. In this calculation the
generalized gradient approximation and the exchange-correlation
functional of Perdew, Burke, and Ernzerhof were applied
\cite{Perdew}. The pseudopotentials of Troullier and Martins
\cite{Troullier} generated by the Fritz Haber Institute code
\cite{FHI} were used; these pseudopotentials were taken from the
ABINIT web page \cite{Abinit}. They were tested by doing calculations
for the bulk $\alpha$-quartz, and these results were compared with
experiments \cite{Levien}. It was found that the computationally
optimized structural parameters of quartz were very close to the
experimental ones; the differences are below $0.1 \%$. Several
properties of silica nanowires were also calculated using the local
density approximation (LDA) with the Teter extended norm-conserving
pseudopotentials  taken from the ABINIT web page
\cite{Teter,Abinit}. The results obtained  using the Teter
pseudopotentials were compared with experiments for bulk
quartz structure, and differences of $0.1 \%$ have been obtained.
Only minor quantitative differences were found between the LDA and GGA results
for silica nanowires. The calculations were performed with
a kinetic-energy cutoff of $35$ Hartree. The wires were positioned in
a supercell of side $30$ a.u. along the x and y directions. The
axis of the wires was taken along the z direction, and the periodic
boundary conditions were applied. The Monkhorst-Pack method with 15
k-points sampling along the z direction was used in the
integration of the Brillouin zone \cite{Monkhorst}. Structural
relaxation for silica nanowires was carried out by performing a
series of self-consistent calculations and computing the forces on
atoms. The geometry optimizations were performed using the Broyden
method of minimization until the forces were less than $2.6  \times
10^{-4}$ eV/\AA. All atoms were allowed to relax without any imposed
constraint.

Infinite Si-O chains were investigated. Two, four, and six atoms in a unit cell of a chain were studied to
explore a possible dimerization and the existence of a zigzag structure.
In previous studies of silica clusters
it has been found that in stable structures there often exists a unit of
two Si$_2$O$_2$ rhombuses sharing one silicon atom. This unit
contains a tetrahedrally bonded Si atom and therefore shows the
structural feature most often present in  the bulk of SiO$_2$. Two
adjacent rhombohedral rings in clusters are perpendicular to each
other. It was calculated in this work that an optimized infinite silica
wire forms if a Si$_4$O$_8$ unit is repeated periodically along a
direction where the silicon atoms are positioned. The Si$_4$O$_8$
unit contains three whole Si$_2$O$_2$ rhombuses.
Infinite tubular
nanostructures of silica,  similar to the finite MgO nanotubes
studied recently \cite{Goranka}, are not stable because their oxygen
atoms are in the 4-fold coordinated configurations. However,
calculations on silica clusters
have shown that the oxygen atom prefers a lower coordination.
In experimental studies of silica nanowires, much bigger structures
having diameters $15-100$ nm and lengths up to tens of millimeters
have been prepared \cite{JLGole,DPYu,MZhang,Mazur}. It has been
shown that these silica nanostructures synthesized in the
laboratories are amorphous.
DFT-based
studies of such already fabricated silica nanowires are not feasible within 
current computational power.

\section{Results and discussion}
\label{sec:3}

\begin{figure}
\center{
\includegraphics{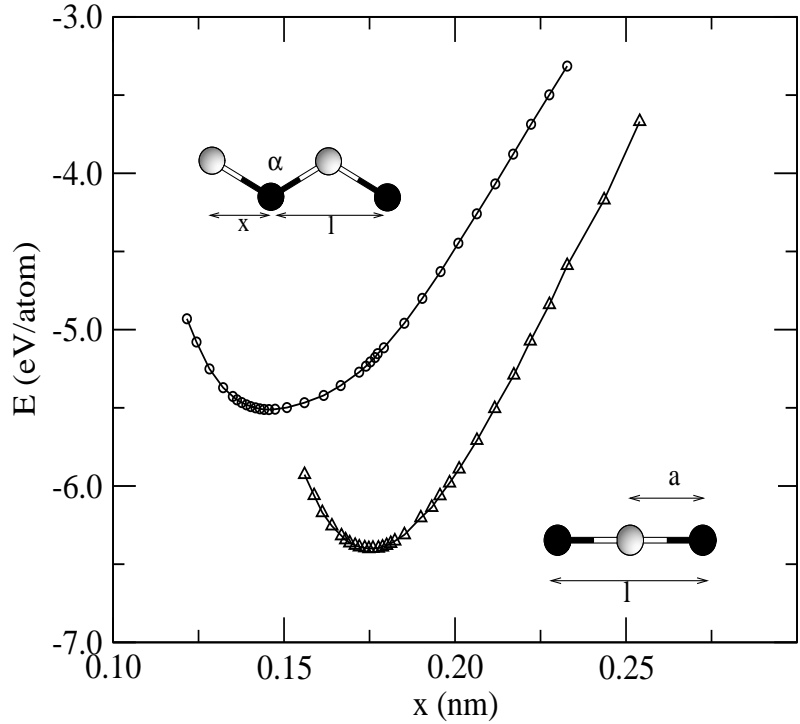}}
\caption{The binding energy  as a function of $x = a$ for a linear chain
(triangles), and $x= a sin(\alpha /2)$ for a zigzag chain (circles), where $a$ is the Si-O distance.
The insets show
the optimized geometries of the chains. Light and dark balls are used to represent
the O and Si atoms respectively. These visualizations were performed using the Rasmol package
\cite{Sayle,Bernstein}.}
\label{fig:fig1}
\end{figure}

\begin{figure}
\center{
\includegraphics*[scale=1.0]{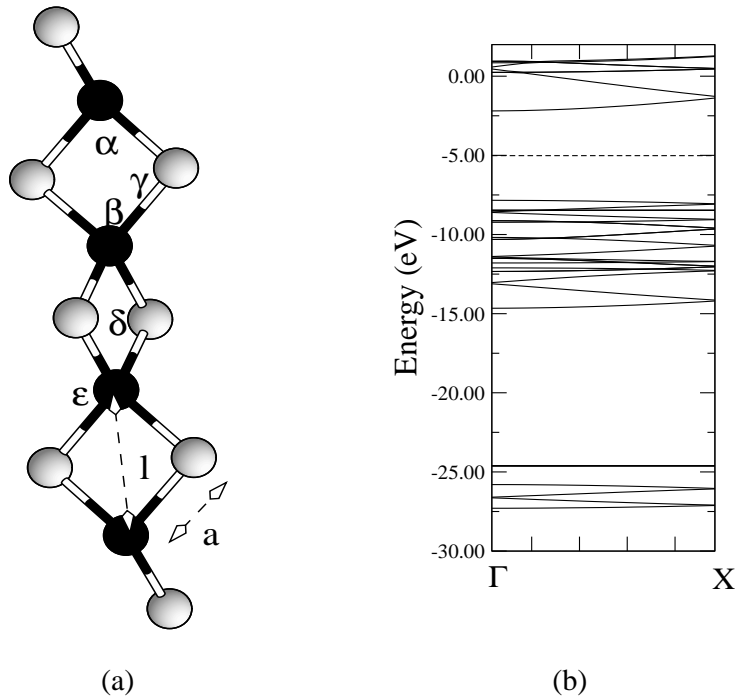}}
\caption{(a) The three-dimensional structural Si$_4$O$_8$ unit which repeats periodically along
the wire axis and forms an optimized but almost unstable infinite nanowire
(angles and lengths are given in the text).
Light and dark balls are used to represent the O and Si atoms respectively.
The middle rhombohedral ring is in the plane perpendicular to two edge rhombuses.
(b) Electronic structure of the structure shown in (a).
The Fermi level is denoted by the dashed line.}
\label{fig:fig2}
\end{figure}

The optimized distances and the binding energies of all nanowires
are presented in table~\ref{tab:table1}. The optimized geometries
of silica chains are shown in figure~\ref{fig:fig1}. No dimerization was found for
the linear chain. The zigzag chain is also stable and its energy is
above that of the linear chain. Nonlinearity of the O-Si-O bonds is
less favorable in a situation where there is no additional oxygen
atoms, as in the case of the bulk tetrahedral SiO$_4$ bonding. The
optimized structure of the Si$_4$O$_8$ unit is shown in
figure~\ref{fig:fig2}(a). It is well known that the Si-O distance in
the bulk silica  is most often about $0.16$ nm. It was calculated
here that a larger Si-O distance of $0.175$ nm exists in a linear
chain, $0.170$ nm in zigzag one, and $0.167$ nm in a Si$_4$O$_8$
nanowire. In the zigzag chain the angles are $\alpha = 118^{\circ}$.
The width of the nanowire shown in figure~\ref{fig:fig2}(a) is up to
about $0.24$ nm. In the Si$_4$O$_8$ wire the oxygen atoms are bonded
to two silicon atoms and the silicon atoms are bonded to four oxygen
atoms. Such SiO$_4$ tetrahedra are typical for bulk materials
involving silicon and oxygen. In the rhombuses of the  Si$_4$O$_8$
wire the Si-O-Si angles are $\delta =89.4^{\circ}$ and $\gamma
=89.5^{\circ}$, while the  O-Si-O ones are $\beta = 90.4^{\circ}$
and $\alpha = 90.6^{\circ}$. The O-Si-O angle is $\epsilon =
119.7^{\circ}$ when the oxygen atoms are in adjacent rhombuses.
Thus, the coordination of the silicon atoms is distorted from
an ideal tetrahedral geometry.
\begin{table}
\caption{\label{tab:table1}The geometry and the binding energy of optimized structures.
In this table a is the Si-O distance, while l is
the Si-Si distance.
The length unit is nm. Energies E are given in eV/atom.}
\begin{indented}
\item[]\begin{tabular}{cccc}
\br
Structure& Linear chain &Zigzag chain & Si$_4$O$_8$ nanowire\\
\mr
a &0.175 & 0.170 &0.167\\
l &0.35  & 0.291 &0.234; 0.235\\
E &-6.40 &-5.51 & -7.38 \\
\br
\end{tabular}
\end{indented}
\end{table}

Figure~\ref{fig:fig1}  also presents the bonding wells for the chains. The minima
are rather pronounced and show a substantial stability of these
nanowires. By contrast, it was not possible to obtain a similar
figure for the Si$_4$O$_8$ wire. Even very small perturbations
($\sim 1\%$) of the length along the wire axis destabilize the
Si$_4$O$_8$ wire. A small difference between the angle within one rhombus
exists ($\beta = 90.4^{\circ}$ vs $\alpha = 90.6^{\circ}$).
It was not possible to stabilize such a
three-dimensional thin wire using a smaller Si$_2$O$_2$ cell.
The Si$_4$O$_8$ nanowire is at the border of instability.
However, it was also found that
the calculation where the LDA approximation to the
DFT theory with the Teter extended norm-conserving pseudopotentials
\cite{Teter} was used produces a similar optimized Si$_4$O$_8$ infinite wire.
For example, in this LDA approximation the Si-Si distance is $l=0.232$ nm, whereas the Si-O distances
are $a=0.164$; $0.165$ nm.
It  should be
possible to assemble silica chains on the surfaces, using various
nanotubes and nanowires, or long channels in porous materials. The
role of the substrate is to increase the stability of very thin
silica nanowires.

\begin{figure}
\center{
\includegraphics*[scale=1.0]{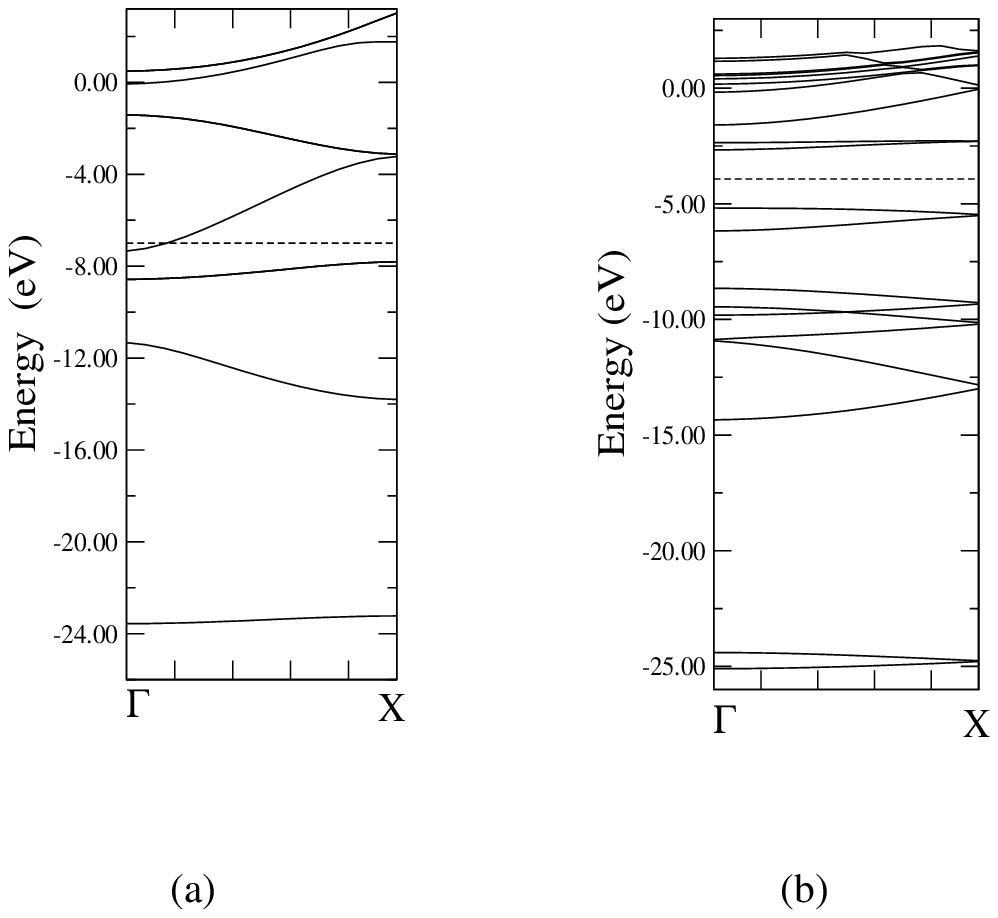}}
\caption{Energy band structures: (a) linear chain, (b) zigzag chain.
The Fermi level is denoted by the dashed line.}
\label{fig:fig3}
\end{figure}

The band structure of silica chains is
shown in figure~\ref{fig:fig3}.
The plot of the electronic structure of a linear chain (presented in
figure~\ref{fig:fig3}(a)) shows that one band  crosses the Fermi
level; therefore this system is metallic. The electronic structure
of a zigzag chain is shown in figure~\ref{fig:fig3}(b). This wire
is an insulator. The band plot in figure~\ref{fig:fig2}(b) shows that
the Si$_4$O$_8$ nanowire is also an insulator. When the number of
neighbours in Si-O nanowires increases, electronic behavior goes
from metallic to insulating, as in the bulk. At the Gamma point, the
difference between the valence and conduction band is $5.646$ eV for
the  Si$_4$O$_8$ nanowire, and $2.514$ eV for the zigzag chain.
Tetrahedral SiO$_4$ clusters exist in the Si$_4$O$_8$ nanowire. This
structure is similar to the fragments of the cristobalite bulk
lattice. The three-dimensional Si$_4$O$_8$ wire behaves as an insulator,
and a similar electronic behavior and a band gap value exist in
the cristobalite crystal \cite{YXu}. Table 1 shows that the Si-O
and Si-Si distances are smaller in the zigzag chain than in the linear
one. This compression of bonds as a result of the rearrangement of
atoms into the zigzag chain removes a crossing band from the Fermi
level, and an insulating behavior arises in this structure. The Si-O
distance in the linear chain is larger than in the majority of silica
bulk phases, as well as in the zigzag and Si$_4$O$_8$ wires. That decreases
the extent of $\pi$ bonding between silicon and oxygen atoms in the linear wire. 
The weak metallic behavior arises in the linear silica chain as a
consequence of this weaker bonding and a small coordination.

Atomic charges were calculated using the Hirshfeld partitioning of
the electron density \cite{Hirshfeld,deProft,Nalewajski}.
The Hirshfeld method (or ``stockholder'' partitioning) uses the charge density distribution to determine
atomic charges in the molecule or nanostructure.
First, the reference state of the promolecule density is defined as
$\rho ^{pro} (\vec r) = \Sigma_A {\rho_A (\vec r)},$
where $\rho_A (\vec r)$ is the electron density of the isolated atom A placed at its position in the molecule.
The atomic charge is
\begin{equation}
q_A = - \int {\delta \rho _A (\vec r) d \vec r},
\end{equation}
where $\delta \rho _A (\vec r)$ is the atomic deformation density given by
\begin{equation}
\delta \rho _A (\vec r) = w_A (\vec r) \Delta \rho(\vec r).
\end{equation}
In equation (2), $w_A(\vec r)$ is the relative contribution (``share'') of the atom A in the promolecule,
whereas $\Delta \rho(\vec r)$ is the molecular deformation density.
The sharing factor is a weight that determines a
relative contribution of the atom $x$ to the promolecule
density in the point $r$. It  is defined as
\begin{equation}
w_A (\vec r) = \frac {\rho _A (\vec r)}{\rho ^{pro}(\vec r)}.
\end{equation}
The molecular deformation density (used in equation (2)) is
\begin{equation}
\Delta (\vec r) = \rho(\vec r) - \rho^{pro}(\vec r),
\end{equation}
where $\rho(\vec r)$ is the molecular electron density.
The Hirshfeld partitioning is almost insensitive to the basis set and  minimizes
missing information \cite{Hirshfeld,deProft,Nalewajski}. The
Hirshfeld charges are presented in table~\ref{tab:table2}. The
calculations show that for all silica wires the charge transfer
occurs from Si to O atoms. This indicates ionic bonding. All oxygen atoms
get similar amounts of the electron density,
regardless of the structure.
\begin{table}
\caption{\label{tab:table2}
The Hirshfeld atomic charges. The calculated average charge transfers $\delta Q$ are shown.}
\begin{indented}
\item[]\begin{tabular}{cccc}
\br
Structure      & Linear chain   & Zigzag chain  &Si$_4$O$_8$  nanowire \\
\mr
$\delta Q(Si)$ &   0.212   &0.275 & 0.446 \\
$\delta Q(O)$  &   -0.212  &-0.275 &-0.225  \\
\br
\end{tabular}
\end{indented}
\end{table}

\begin{figure}
\center{\includegraphics*[scale=0.8]{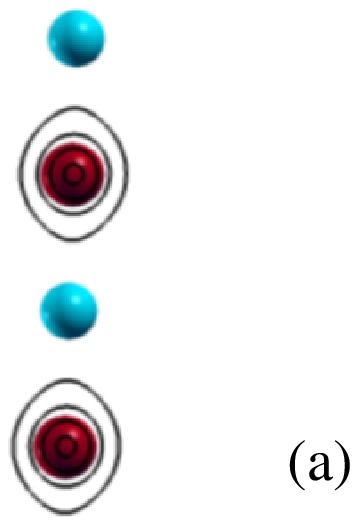}
\includegraphics*[scale=0.8]{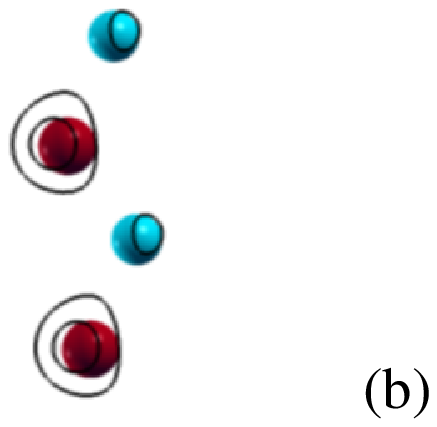}
\includegraphics*[scale=0.8]{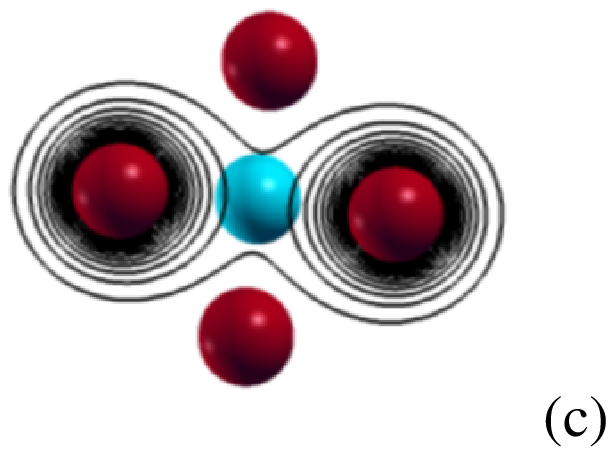}}
\caption{Charge density contour plots drawn using 0.4
e/$a_0^3$ isovalues. The dark spheres are O, whereas the light ones are
Si atoms. Side views of unit cells for: (a) linear chain, (b) zigzag
chain. (c) A top view of the cell for the Si$_4$O$_8$ nanowire:
isolines around two top oxygen atoms are shown and the central ball
represents the strand of Si atoms. } \label{fig:fig4}
\end{figure}

The character of the bonding was also analysed using the electronic charge density.
In figure \ref{fig:fig4} the charge density isosurface plots are presented.
This visualization was performed by the XCrySDen package
\cite{Tone}.
The well-defined spherical charges are located and accumulated on the oxygen atoms.
Similar charge density plots that show a predominantly ionic bonding
have been, for example, obtained for bulk $\alpha$-quartz
\cite{Chelikowsky1}.

\section{Conclusions}
\label{sec:4}

Three configurations of infinite silica nanowires were optimized and
studied using \textit{ab initio} DFT calculations in the
pseudopotential approximation. The structural properties of these wires
were investigated. It was found that a linear chain  is energetically
more favorable than a zigzag wire. The calculations of the bonding
wells showed that both chains are stable, whereas the infinite
Si$_4$O$_8$ wire is at the border of instability. The Hirshfeld
charges were calculated and the results show that a similar transfer
of a charge to oxygen atoms exists for all wires. It was found that
the zigzag chain and the Si$_4$O$_8$ wire are insulators, while a
single state crosses the Fermi level in the band plot of the linear
chain. The existence of a metallic state offers the possibility to use
simple long silica chains in conducting nanodevices without doping.
It is possible to deposit and assemble these chains on  various
surfaces, nanotubes, or inside the long and wide pores of suitable
bulk materials.

\ack
{This work has been carried under the HR-MZOS project No. 0119255
\textquotedblleft Dynamical Properties and Spectroscopy of Surfaces
and Nanostructures\textquotedblright. Part of the calculations was
done on the cluster of PCs at the University Computing Center
SRCE, Zagreb.}

\end{document}